\documentclass[11pt]{article}
\usepackage{fullpage}
\usepackage{graphicx}

\newcommand{\search}
    {\ensuremath{\mathsf{search}}}
\newcommand{\lookup}
    {\ensuremath{\mathsf{lookup}}}

\newcommand{\no}[1]{}

\begin{document}

\title{Practical Top-K Document Retrieval in Reduced Space
    \thanks{
    Partially funded by the Millennium Institute for Cell Dynamics and 
    Biotechnology (ICDB), Grant ICM P05-001-F, Mideplan, Chile; and by 
    Fondecyt Grant 1-110066, Chile.}
    }
\author{Gonzalo Navarro ~~~~~~~~~~~~~~~~~~~~~~~~~ Daniel Valenzuela \\
\ \\
\small
Dept. of Computer Science,
    University of Chile, {\tt \{gnavarro,dvalenzu\}@dcc.uchile.cl}
}

\date{}

\maketitle

\begin{abstract}
Supporting top-$k$ document retrieval queries on general text databases, that 
is, finding the $k$ documents where a given pattern occurs most frequently, has
become a topic of interest with practical applications. 
While the problem has been solved in 
optimal time and linear space, the actual space usage is a serious concern. In 
this paper we study various reduced-space structures that support top-$k$ 
retrieval and propose new alternatives. Our experimental results show 
that our novel algorithms and data structures dominate almost all the 
space/time tradeoff.
\end{abstract}

\newpage

\section{Introduction} \label{sec:introduction}

Ranked document retrieval is the basic task of most search engines. It
consists in preprocessing a collection of $d$ {\em documents}, 
$\mathcal{D} = \{ D_1, D_2, \ldots, D_d\}$, so that later, given a {\em query
pattern} $P$ and a {\em threshold} $k$, one quickly finds the $k$ documents
where $P$ is ``most relevant''.

The best known application scenario is that of documents being formed by 
natural language texts, that is, sequences of words, and the query patterns 
being words,
phrases (sequences of words), or sets of words or phrases. Several relevance
measures are used, which attempt to establish the significance of the query in a
given document \cite{BYRN11}. The {\em term frequency}, that is, the number of 
times the pattern appears in the document, is the main component of most
measures.

Ranked document retrieval is usually solved with some variant of a simple 
structure called an inverted index \cite{WMB99,BYRN11}. This structure, which
is behind most search engines, handles well natural language collections. 
However, the term ``natural language'' hides several assumptions that are 
key to the efficiency of that solution: the text must be easily tokenized into 
a sequence of words, there must not be too many different words, and queries 
must be whole words or phrases. 

Those assumptions do not hold in various applications where document retrieval 
is of interest. The most obvious ones are documents written in Oriental 
languages such as Chinese or Korean, where it is not easy to split words 
automatically, and search engines treat the text as a sequence of symbols, 
so that queries can retrieve any substring of the text. 
Other applications
simply do not have a concept of word, yet ranked retrieval would be of
interest: DNA or protein sequence databases where one seeks the sequences
where a short marker appears frequently, source code repositories where one
looks for functions making heavy use of an expression or function call, MIDI 
sequence databases where one seeks for pieces where a given short passage is
repeated, 
and so on.

These problems are modeled as a text collection where the documents $D_i$ are
strings over an alphabet $\Sigma$, of size $\sigma$, and the queries are also
simple strings. The most
popular relevance measure is the plain term frequency, that is, the number of
occurrences of the string $P$ in the strings $D_i$.\footnote{It is usual to
combine the term frequency with the so-called ``inverse document frequency'', 
but this makes a difference only in the more complex bag-of-word queries,
which have not yet been addressed in this context.} We call 
$n = \sum |D_i|$ the collection size and $m=|P|$ the pattern length.

Muthukrishnan \cite{Mut02} pioneered the research on document retrieval for
general strings. He solved the simpler problem of ``document listing'':
report the $occ$ distinct documents where $P$ appears in optimal time 
$O(m+occ)$ and linear space, $O(n)$ integers (or $O(n\log n)$ bits). 
Muthukrishnan also considered various other document retrieval problems, 
but not top-$k$ retrieval.

The first efficient solution for the top-$k$ retrieval problem was introduced
by Hon, Shah, and Wu \cite{HSW09}. They achieved $O(m+\log n \log\log n + k)$
time, yet the space was superlinear, $O(n\log^2 n)$ bits. Soon, Hon, Shah, and
Vitter \cite{HSV09} achieved $O(m+k\log k)$ time and linear space, $O(n\log n)$
bits. Recently, Navarro and Nekrich \cite{NN12} achieved optimal time,
$O(m+k)$, and reduced the space from $O(n\log n)$ to 
$O(n(\log\sigma+\log d))$ bits (albeit the constant is not small). 

While these solutions seem to close the problem, it turns out that the space
required by $O(n\log n)$-bit solutions is way excessive for practical 
applications. A recent space-conscious implementation of Hon et al.'s index 
\cite{PTSVC11} showed that it requires at least 5 times the text size. 

Motivated by this challenge, there has been a parallel research track on how
to reduce the space of these solutions, while retaining efficient search time
\cite{Sad07,VM07,HSV09,GNP10,CNPT10,BN11,NPV11,HST11}. In this work we 
introduce a new variant with relevant theoretical and practical properties,
and show experimentally that it dominates previous work. The next
section puts our contribution in context.

\section{Related Work}

Most of the data structures for general text searching, and in particular
the classical ones for document retrieval \cite{Mut02,HSV09}, build on
on {\em suffix arrays} \cite{MM93} and {\em suffix trees} 
\cite{Wei73,Apo85}. Regard the collection $\mathcal{D}$ as a single text 
$T[1,n] = D_1 D_2 \ldots D_d$, where each $D_i$ is terminated by a special 
symbol ``\$''. A suffix array $A[1,n]$ is a permutation of the values $[1,n]$ 
that points to all the {\em suffixes} of $T$: $A[i]$ points to the suffix
$T[A[i],n]$. The suffixes are {\em lexicographically sorted} in $A$:
$T[A[i],n] < T[A[i{+}1],n]$ for all $1 \le i < n$. Since the occurrences of 
any pattern $P$ in $T$ correspond to suffixes of $T$ that are prefixed by $P$, 
the occurrences are pointed from a contiguous area in the suffix array
$A[sp,ep]$. A simple binary search finds $sp$ and $ep$ in $O(m\log n)$ time
\cite{MM93}. A suffix tree is a digital tree with $O(n)$ nodes where all the 
suffixes of $T$ are inserted and unary paths are compacted. Every internal node
of the suffix tree corresponds to a repeated substring of $T$ and its 
associated suffix array interval;, suffix tree leaves correspond to the 
suffixes and their corresponding suffix array cells. 
A top-down traversal in the suffix tree finds the internal node
(called the {\em locus} of $P$) from where all the suffixes prefixed with $P$ 
descend, in $O(m)$ time. Once $sp$ and $ep$ are known, the top-$k$ query finds 
the $k$ documents where most suffixes in $A[sp,ep]$ start.

A first step towards reducing the space in top-$k$ solutions is to compress
the suffix array. {\em Compressed suffix arrays (CSAs)} simulate
a suffix array within as little as $nH_k(T)+o(n\log\sigma)$ bits, for any 
$k \le \alpha\log_\sigma n$ and any constant $0<\alpha<1$. Here $H_k(T)$ is 
the $k$-th order entropy of $T$ \cite{Man01}, a measure of its statistical
compressibility. The CSA, using $|CSA|$ bits, finds $sp$
and $ep$ in time $\search(m)$, and computes any cell $A[i]$, and even
$A^{-1}[i]$, in time $\lookup(n)$. For example, a CSA achieving the small 
space given above \cite{FMMN07} achieves 
$\search(m)=O(m(1+\frac{\log\sigma}{\log\log n}))$
and $\lookup(n) = O(\log^{1+\epsilon} n)$ for any constant $\epsilon>0$.
CSAs also replace the collection, as they can extract any substring of $T$.

In their very same foundational paper, Hon et al.~\cite{HSV09} proposed an
alternative succinct data structure to solve the top-$k$ problem. Building on
a solution by Sadakane \cite{Sad07} for document listing, they use
a CSA for $T$ and one smaller CSA for each document $D_i$, plus little extra
data, for a total space of $2|CSA|+o(n)+d\log(n/d)+O(d)$ bits. They achieve
time $O(\search(m)+k\log^{3+\epsilon} n \cdot \lookup(n))$, for any constant
$\epsilon>0$. Gagie, Navarro, and Puglisi~\cite{GNP10} slightly reduced the 
time to $O(\search(m)+k\log d \log(d/k)\log^{1+\epsilon} n \cdot \lookup(n))$, 
and Belazzougui and Navarro \cite{BN11} further improved it to
$O(\search(m)+k\log k \log(d/k) \log^\epsilon n \cdot \lookup(n))$.

The essence of the succinct solution by Hon et al.~\cite{HSV09} is to
preprocess top-$k$ answers for the lowest suffix tree nodes containing any
range $A[i\cdot g,j \cdot g]$ for some sampling parameter $g$. Given the query 
interval $A[sp,ep]$, they find the highest preprocessed suffix tree node whose 
interval $[sp',ep']$ is contained in $[sp,ep]$. They show that $sp'-sp < g$ and
$ep-ep' < g$, and then the cost of correcting the precomputed answer using the 
extra occurrences at $A[sp,sp'{-}1]$ and $A[ep'{+}1,ep]$ is bounded. For each 
such extra occurrence 
$A[i]$, one finds out its document, computes the number of occurrences of $P$ 
within that document, and lets the document compete in the top-$k$ precomputed 
list. Hon et al.\ use the individual CSAs and other data structures to carry 
out this task. The subsequent improvements \cite{GNP10,BN11} are due to small 
optimizations on this basic design.

Gagie et al.~\cite{GNP10} also pointed out that in fact Hon et al.'s solution
can run on any other data structure able to (1) telling which is the
document corresponding to a given $A[i]$, and (2) counting how many times does
the same document appear in any interval $A[sp,ep]$. A structure that is
suitable for this task is the {\em document array} $D[1,n]$, where $D[i]$ is
the document $A[i]$ belongs to \cite{Mut02}. While in Hon et al.'s solution 
this is computed from $A[i]$ using $d\log(n/d)+O(d)$ extra bits \cite{Sad07}, 
we need more machinery for task (2). A good alternative was proposed by
M\"akinen and Valim\"aki \cite{VM07} in order to reduce the space of 
Muthukrishnan's document listing solution \cite{Mut02}. The structure is
a {\em wavelet tree} \cite{GGV03} on $D$. The wavelet tree represents $D$
using $n\log d + o(n)\log d$ bits and not only computes any $D[i]$ in 
$O(\log d)$ time, but it can also compute operation $rank_i(D,j)$, which
is the number of occurrences of document $i$ in $D[1,j]$, within the same 
time. This solves operation (2) as $rank_{D[i]}(D,ep)-rank_{D[i]}(D,sp{-}1)$.
With the obvious disadvantage of the considerable extra space to represent
$D$, this solution changes $\lookup(n)$ by $\log d$ in the query time. Gagie 
et al.\ show many other combinations that solve (1) and (2). One of the fastest
uses Golynski et al.'s representation \cite{GMR06} on $D$ and, within the same 
space, changes $\lookup(n)$ to $\log\log d$ in the time. Very recently, Hon, 
Shah, and Thankachan \cite{HST11} presented new combinations in the line of
Gagie et al., using also faster CSAs. The least space-consuming one requires
$n\log d + n\,o(\log d)$ bits of extra space on top of the CSA of $T$, and 
improves the time to $O(\search(m)+k(\log k + (\log\log n)^{2+\epsilon}))$.

Belazzougui and Navarro \cite{BN11} used an approach based on minimum perfect
hash functions to replace the array $D$ by a weaker data structure that takes
$O(n \log\log\log d)$ bits of space and supports the search in time
$O(\search(m) + k \log k \log^{1+\epsilon} n \cdot \lookup(n))$. This is 
solution is intermediate between representing $D$ or the individual CSAs
and it could have practical relevance.

Culpepper, Navarro, Puglisi, and Turpin~\cite{CNPT10} built on an improved 
document listing algorithm on wavelet trees \cite{GPT09} to achieve two top-$k$ 
algorithms, called {\em Quantile} and {\em Greedy}, that use the wavelet tree 
alone (i.e., without Hon et al.'s~\cite{HSV09} extra structures). Despite their
worst-case complexity being as bad as extracting one by one the results in 
$A[sp,ep]$, that is, $O((ep-sp+1)\log d)$, in practice the algorithms performed
very well, being Greedy superior. They implemented
Sadakane's solution \cite{Sad07} of using individual CSAs for the documents
and showed that the overheads are very high in practice. Navarro, Puglisi,
and Valenzuela~\cite{NPV11} arrived at the same conclusion, showing that Hon et 
al.'s original succinct scheme is not promising in practice: both space and 
time were much higher in practice than Culpepper et al.'s solution. However, 
their preliminary experiments \cite{NPV11} showed that Hon et al.'s scheme 
could compete when running on wavelet trees.

Navarro et al.~\cite{NPV11} also presented the first implemented alternative to
reduce the space of wavelet trees, by using Re-Pair compression \cite{LM00}
on the bitmaps. They showed that significant reductions in space were possible
in exchange for an increase in the response time of Culpepper et al.'s Greedy
algorithm (half the space and twice the time is a common figure). 

This review exposes interesting contrasts between the theory and the practice in
this area. On one hand, the structures that are in theory larger and faster
(i.e., the $n\log d$-bits wavelet tree versus a second CSA of at most
$n\log\sigma$ bits) are in practice {\em smaller} and faster. On the other
hand, algorithms with no worst-case bound (Culpepper et al.'s \cite{CNPT10}) 
perform very well in practice. Yet, the space of wavelet trees is still 
considerably large in practice (about twice the plain size of $T$ in several
test collections \cite{NPV11}), especially if we realize that they represent 
totally redundant information that could be extracted from the CSA of $T$.

In this paper we study a new practical alternative. We use Hon et al.'s 
\cite{HSV09} succinct structure on top of a wavelet tree, but instead of brute 
force we use a variant of Culpepper et al.'s \cite{CNPT10} method to find the 
extra candidate documents in $A[sp,sp'{-}1]$ and $A[ep'{+}1,ep]$. We can regard 
the combination as either method boosting the other. Culpepper et al.\ boost Hon
et al.'s method, while retaining its good worst-case complexities, as they 
find the extra occurrences more cleverly than by enumerating them all. Hon et 
al.\ boost plain Culpepper et al.'s method by having precomputed a large part 
of the range, and thus ensuring that only small intervals have to be handled.

We consider the plain and the compressed wavelet tree representations,
and the straightforward and novel representations of Hon et al.'s succinct 
structure. We compare
these alternatives with the original Culpepper et al.'s method (on plain and
compressed wavelet trees), to test the hypothesis that adding Hon et al.'s 
structure is worth the extra space. Similarly, we include in the comparison
the basic Hon et al.'s method (with their structure compressed or not) over
Golynski et al.'s \cite{GMR06} sequence representation, 
to test the hypothesis that using Culpepper et al.'s method
over the wavelet tree is worth compared to the brute force method over
the fastest sequence representation \cite{GMR06}. This brute force method is
also at the core of the new proposal by Hon et al. \cite{HST11}.

Our experiments show that our new algorithms and data structures dominate
almost all the space/time tradeoff for this problem, becoming a new practical
reference point.

\section{Implementing Hon et al.'s Succinct Structure}
\label{sec:SSGST}

The succinct structure of Hon et al.~\cite{HSV09} is a sparse generalized
suffix tree of $T$ (SGST; ``generalized'' means it indexes $d$ strings). It is
obtained by cutting $A[1,n]$ into blocks of length $g$ and sampling the first
and last cell of each block (recall that cells of $A$ are leaves of the suffix
tree). Then all the lowest common ancestors ({\em lca}) of pairs of sampled 
leaves are marked, and a tree $\tau_k$ is formed with those (at most) $2n/g$ 
marked internal nodes. The top-$k$ answer is stored for each marked node, using 
$O((n/g)k\log n)$ bits. This is done for $k=1,2,4,\ldots$, and
parameter $g$ is of the form $g=k \cdot g'$. The final space is 
$O((n/g')\log d \log n)$ bits. This is made $o(n)$ by properly choosing $g'$. 

To answer top-$k$ queries, they search the CSA for $P$, to obtain 
the suffix range $A[sp,ep]$ of the pattern. Then they turn to the closest 
higher power of two of $k$, $k^*=2^{\lceil \log k\rceil}$, and let 
$g = k^*\cdot g'$ be the corresponding $g$ value. They now find the locus of 
$P$ in the tree $\tau_{k^*}$ by descending from the root until finding the
first node $v$ whose interval $[sp_v,ep_v]$ is contained in $[sp,ep]$.
They have at $v$ the top-$k$ candidates for $[sp_v,ep_v]$ and 
have to correct the answer considering $[sp,sp_v{-}1]$ and $[ep_v{+}1,ep]$.
Now we introduce two implementations of this idea.

\subsection{Sparsified Generalized Suffix Tree (SGST) }

Let us call $l_i = A[i]$ the $i$-th leaf. Given a value of $k$ we define
$g = k \cdot g'$, for a space/time tradeoff parameter $g'$, and sample 
$n/g$ leaves $l_1, l_{g+1}, l_{2g+1}, \ldots$, instead of sampling $2n/g$ 
leaves as in the theoretical proposal. We mark internal SGST
nodes $lca(l_1,l_{g+1}), lca(l_{g+1},l_{2g+1}), \ldots$. It is not hard to 
prove that any $v=lca(l_{ig+1},l_{jg+1})$ is also $v=lca(l_{rg+1},l_{(r+1)g+1})$
for some $r$ (more precisely, $r$ is the rightmost sampled leaf descending from
the child of $v$ that is an ancestor of $l_{ig+1}$). Therefore these $n/g$
SGST nodes are sufficient and can be computed in linear time \cite{BFC00}.

Now we note that there is a great deal of redundancy in the $\log d$ trees 
$\tau_k$, since the nodes of $\tau_{2k}$ are included in those
of $\tau_k$, and the $2k$ candidates stored in the nodes of $\tau_{2k}$
contain those in the corresponding nodes of $\tau_k$. 
To factor out some of this redundancy we store only one tree $\tau$, whose
nodes are the same of $\tau_1$, and record the {\em class} $c(v)$ of each node 
$v \in \tau$. This is $c(v) = \max \{ k,~v \in \tau_k \}$ and can be stored in 
$\log\log d$ bits. Each node $v \in \tau$ stores the top-$c(v)$ candidates
corresponding to its interval, using $c(v)\log d$ bits, and their frequencies,
using $c(v)\log n$ bits, plus a pointer to the table\footnote{Actually, an 
index to a big table where all these small tables are stored consecutively.}, 
and the interval itself, $[sp_v,ep_v]$, using $2\log n$ bits. 
All the information on intervals and candidates is factored in this way, 
saving space. Note that the class does not necessarily decrease monotonically 
in a root-to-leaf path of $\tau$, thus we store all the topologies 
independently to allow for efficient traversal of the $\tau_k$ trees, for 
$k>1$. Apart from topology information, each node of such $\tau_k$ trees
contains just a pointer to the corresponding node in $\tau$, using 
$\log|\tau|$ bits. 

In our first data structure, the topology of the trees $\tau$ and $\tau_k$ is 
represented using pointers of $\log |\tau|$ and $\log |\tau_k|$ bits,
respectively. To answer top-$k$ queries, we find the range $A[sp,ep]$ using a 
CSA (whose space and negligible time will not be reported because it is 
orthogonal to all the data structures). Now we find the locus in the 
appropriate tree $\tau_{k^*}$ top-down, binary searching the intervals 
$[sp_v,ep_v]$ of the children of the current node, and
extracting those intervals using the pointer to $\tau$. By the properties of 
the sampling \cite{HSV09} it follows that we will traverse in this descent 
nodes $v \in \tau_{k^*}$ such that $[sp,ep] \subseteq [sp_v,ep_v]$, until 
reaching a node $v$ so that $[sp_v,ep_v] = [sp',ep'] \subseteq [sp,ep] 
\subseteq [sp'-g,ep'+g]$ (or reaching a leaf $u \in \tau_k$ such that
$[sp,ep] \subseteq [sp_u,ep_u]$, in which case $ep-sp+1 < 2g$). 
This $v$ is the locus of $P$ in $\tau_{k^*}$, and we 
find it in time $O(m\log\sigma)$. This time is negligible compared to the 
subsequent costs, as well as is the search using the CSA.

\subsection{Succinct SGST}

Our second implementation uses a pointerless representation of the tree
topologies. Although the tree operations are slightly slower than on a 
pointer-based representation, this slowdown occurs on a not too significant
part of the search process, and a succinct representation allows one to reduce 
the sampling parameter $g$ for the same space usage.

Arroyuelo et al.~\cite{ACNS10} showed that, for the functionality
it provides, the most promising succinct representation of trees 
is the so-called Level-Order Unary Degree Sequence (LOUDS) \cite{Jac89}. 
It requires $2N+o(N)$ bits of space (in practice, as little as $2.1\,N$) 
to represent a tree of $N$ nodes, and it solves many operations in 
constant time (less than a microsecond in practice). 

We use that implementation \cite{ACNS10}.
The shape of the tree is stored using a single binary sequence, as follows.
Starting with an empty bitstring, every node is visited in level
order starting from the root. Each node with $c$ children is encoded by
writing its arity in unary, that is, $1^c0$ is appended to the bitstring. 
Each node is identified with the position in the bitstring where the
encoding of the node begins. We store the values $sp_v$ and $ep_v$ in a
separate array, indexed by the position of the node $v$ in the bitstring.
Other node data such as pointers to $\tau$ (in $\tau_k$) and to the candidates
(in $\tau$) are stored in the same way.
The space can be further reduced by storing only the identifiers of the 
candidates, and their frequencies are computed on the fly using $rank$ on
the wavelet tree of $D$. 

\section{A New Top-$k$ Algorithm}
\label{sec:ALGORITHMS}

We run a combination of the algorithm by Hon et al.~\cite{HSV09} and those of
Culpepper et al.~\cite{CNPT10}, over a wavelet tree representation of the
document array $D[1,n]$. Culpepper et al.\ introduce, among others, a document 
listing method (DFS) and a Greedy top-$k$ heuristic. We adapt these to our 
particular top-$k$ subproblem.

If the search for the locus of $P$ ends at a leaf $u$ that still contains the
interval $[sp,ep]$, Hon et al.\ simply scan $A[sp,ep]$ by brute force and
accumulate frequencies. We use instead Culpepper et al.'s Greedy algorithm 
which is always better than a brute-force scanning.

When, instead, the locus of $P$ is a node $v$ where $[sp_v,ep_v] = [sp',ep']
\subseteq [sp,ep]$, we start with the precomputed answer of the $k \le k^*$
most frequent documents in $[sp',ep']$, and update it to consider the 
subintervals $[sp,sp'{-}1]$ and $[ep'{+}1,ep]$. We use the wavelet tree of $D$
to solve the following problem:
Given an interval $D[l,r]$, and two
subintervals $[l_1,r_1]$ and $[l_2,r_2]$, enumerate all
the distinct values in $[l_1,r_1] \cup [l_2,r_2]$ together with
their frequencies in $[l,r]$.
We propose two solutions, which can be seen as generalizations
of heuristics proposed by Culpepper et al.\ \cite{CNPT10}. 

\subsection{Restricted Depth-First Search (DFS)}

\begin{figure}[tb]
\begin{center}
\includegraphics[width=8.9cm]{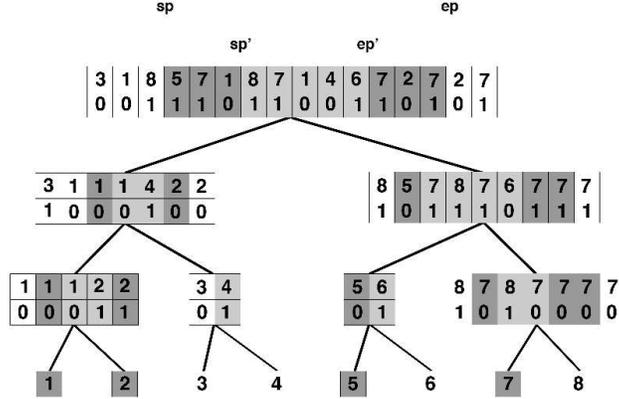}
\vspace*{-1.5cm}
\caption{Restricted DFS to obtain the frequencies of
documents not covered by $\tau_k$. Shaded regions show the
interval $[sp,ep]=[4,14]$ mapped to each wavelet tree node. 
Dark shaded intervals are the projections of the leaves not
covered by $[sp',ep']=[7,11]$.
}
\end{center}
\label{fig:MOD_DFS}
\end{figure}

Figure~\ref{fig:MOD_DFS} illustrates a wavelet tree representation of an
array $D$ (ignore colors for now). At the root, a bitmap 
$B[1,n]$ stores $B[i]=0$ if $D[i] \le d/2$ and $B[i]=1$ otherwise. The left
child of the root is, recursively, a wavelet tree handling the subsequence
of $D$ with values $D[i] \le d/2$, and the right child handles the subsequence
of values $D[i] > d/2$. Only the bitmaps $B$ are actually stored. Added over
the $\log d$ levels, the wavelet tree requires $n\log d$ bits of space. With
$o(n\log d)$ additional bits we answer in constant time any query 
$rank_{0/1}(B,i)$ over any bitmap $B$ \cite{Jac89}.

Note that any interval $D[i,j]$ can be projected into the left child of the
root as
$[i_0,j_0] = [rank_0(B,i{-}1){+}1,rank_0(B,j)]$, and into its right child as
$[i_1,j_1] = [rank_1(B,i{-}1){+}1,rank_1(B,j)]$, where $B$ is the root bitmap.
Those can then be projected recursively into other wavelet tree nodes.

Our restricted DFS algorithm begins at the root of the wavelet tree and
tracks down the intervals $[l,r] = [sp,ep]$, $[l_1,r_1] = [sp,sp'{-}1]$,
and $[l_2,r_2] = [ep'{+}1,ep]$. More precisely, we count the number of zeros
and ones in $B$ in ranges $[l_1,r_1] \cup [l_2,r_2]$, as well as in $[l,r]$,
using a constant number of \emph{rank} operations on $B$. If there are any 
zeros in $[l_1,r_1]\cup[l_2,r_2]$, we map all the intervals into the left 
child of the node and proceed recursively from this node. Similarly, if there
are any ones in $[l_1,r_1]\cup[l_2,r_2]$, we continue on the right child of
the node. When we reach a wavelet tree leaf we report the corresponding
document, and the frequency is the length of the interval $[l,r]$ at the leaf.
Figure~\ref{fig:MOD_DFS} shows an example where we arrive at the leaves
of documents 1, 2, 5 and 7, reporting frequencies 2, 2, 1 and 4,
respectively.

When solving the problem in the context of top-$k$ retrieval, we can prune
some recursive calls. If, at some node, the size of the local interval $[l,r]$ 
is smaller than our current $k$th candidate to the answer, we stop exploring 
its subtree since it cannot contain competitive documents.

\subsection{Restricted Greedy}

Following the idea described by Culpepper et al., we can not only stop the
traversal when $[l,r]$ is too small, but also prioritize the traversal of 
the nodes by their $[l,r]$ value.

We keep a priority queue where we store the wavelet tree nodes yet to process,
and their intervals $[l,r]$, $[l_1,r_1]$, and $[l_2,r_2]$. 
The priority queue begins with one element, the root. Iteratively, we remove
the element with highest $r{-}l{+}1$ value from the queue. If it is a leaf, we
report it. Else, we project the intervals into its left and right children,
and insert each such children containing nonempty intervals $[l_1,r_1]$ or
$[l_2,r_2]$ into the queue. As soon as the $r{-}l{+}1$ value of the element we
extract from the queue is not larger than the $k$th frequency known at the
moment, we can stop.

\subsection{Heaps for the $k$ Most Frequent Candidates}

Our two algorithms solve the query assuming that we can easily know at each
moment which is the $k$th best candidate known up to now. We use a min-heap
data structure for this purpose. It is loaded with the top-$k$ precomputed
candidates corresponding to the interval $[sp',ep']$. At each point, the
top of the heap gives the $k$th known frequency in constant time. Given that
the previous algorithms stop when they reach a wavelet tree node where 
$r{-}l{+}1$ is not larger than the $k$th known frequency, it follows that each
time the algorithms report a new candidate, this is more frequent than our
$k$th known candidate. Thus we replace the top of our heap with the reported
candidate and reorder the heap (which is always of size $k$, or less
until we find $k$ distinct elements in $D[sp,ep]$). Therefore each candidate
reported costs $O(\log d + \log k)$ time (there are also steps that do not
yield any result, but the overall upper bound is still $O(g(\log d+\log k))$).

A remaining issue is that we can find again, in our DFS or Greedy traversal, 
a node that was in the original top-$k$ list, and thus possibly in the heap.
This means that the document had been inserted with its frequency in
$D[sp',ep']$, but since it appears more times in $D[sp,ep]$, we must now
update its frequency, that is, increase it and restore the min-heap
invariant.
It is not hard to maintain a hash table with forward and backward pointers 
to the heap so that we can track their current positions and replace their
values. However, for the small $k$ values used in practice (say, tens or
at most hundreds), it is more practical to scan the heap for each new
candidate to insert than to maintain all those pointers upon all operations.

\section{Experimental Results}

We test the performance of our implementations of Hon et al.'s succinct 
structure combined with a wavelet tree (as explained, the original proposal
is not competitive in practice \cite{NPV11}). 

We used three test collections of different nature:
{\bf ClueWiki} is a 141 MB sample of {\em ClueWeb09}, formed by 3,334 Web 
pages from the English Wikipedia; 
{\bf KGS} is a 75 MB collection of 18,838 sgf-formatted Go game records 
({\tt http://www.u-go.net/gamerecords}); and 
{\bf Proteins} is a 60 MB collection of 143,244 sequences of Human and 
Mouse proteins ({\tt http://www.ebi.ac.uk/swissprot}). 

Our tests were run on a 4-core 8-processors Intel Xeon, 2Ghz each, with 16GB 
RAM and 2MB cache. We compiled using {\tt g++} with full optimization.
For queries, we selected 1,000 substrings at random positions, of length 3 
and 8, and retrieved the top-$k$ documents for each, for $k=1$ and $10$. 

\subsection{Choosing Our Best Variant}

Our first round of experiments compares our different implementations
of SGSTs (i.e., the trees $\tau_k$, see Section~\ref{sec:SSGST}) over a single 
implementation of wavelet tree ({\tt Alpha}, choosing the best value for 
$\alpha$ in each case \cite{NPV11}).
We tested a pointer-based representation of the SGST ({\tt Ptrs}, the
original proposal \cite{HSV09}), 
a LOUDS-based representation ({\tt LOUDS}), 
our variant of {\tt LOUDS} that stores the topologies in a unique tree
$\tau$ ({\tt LIGHT}),
and our variant of {\tt LIGHT} that does not store frequencies of the top-$k$
candidates ({\tt XLIGHT}).
We consider sampling steps of 200 and 400 for $g'$. For each value of $g$, 
we obtain a curve with various sampling steps for the $rank$
computations on the wavelet tree bitmaps.

We also tested different algorithms to find the top-$k$ among the precomputed
candidates and remaining leaves (see Section~\ref{sec:ALGORITHMS}): 
Our modified greedy ({\tt Greedy}), our modified depth-first-search ({\tt DFS}),
and the brute-force selection procedure of the original proposal \cite{HSV09}
on top of the same wavelet tree ({\tt Select}). As this is orthogonal to the 
data structures used, we compare these algorithms only on top of the {\tt Ptrs}
structure. The other structures will use the best method.

\begin{figure}[p]
\centerline{
\includegraphics[angle=-90,width=0.49\textwidth]{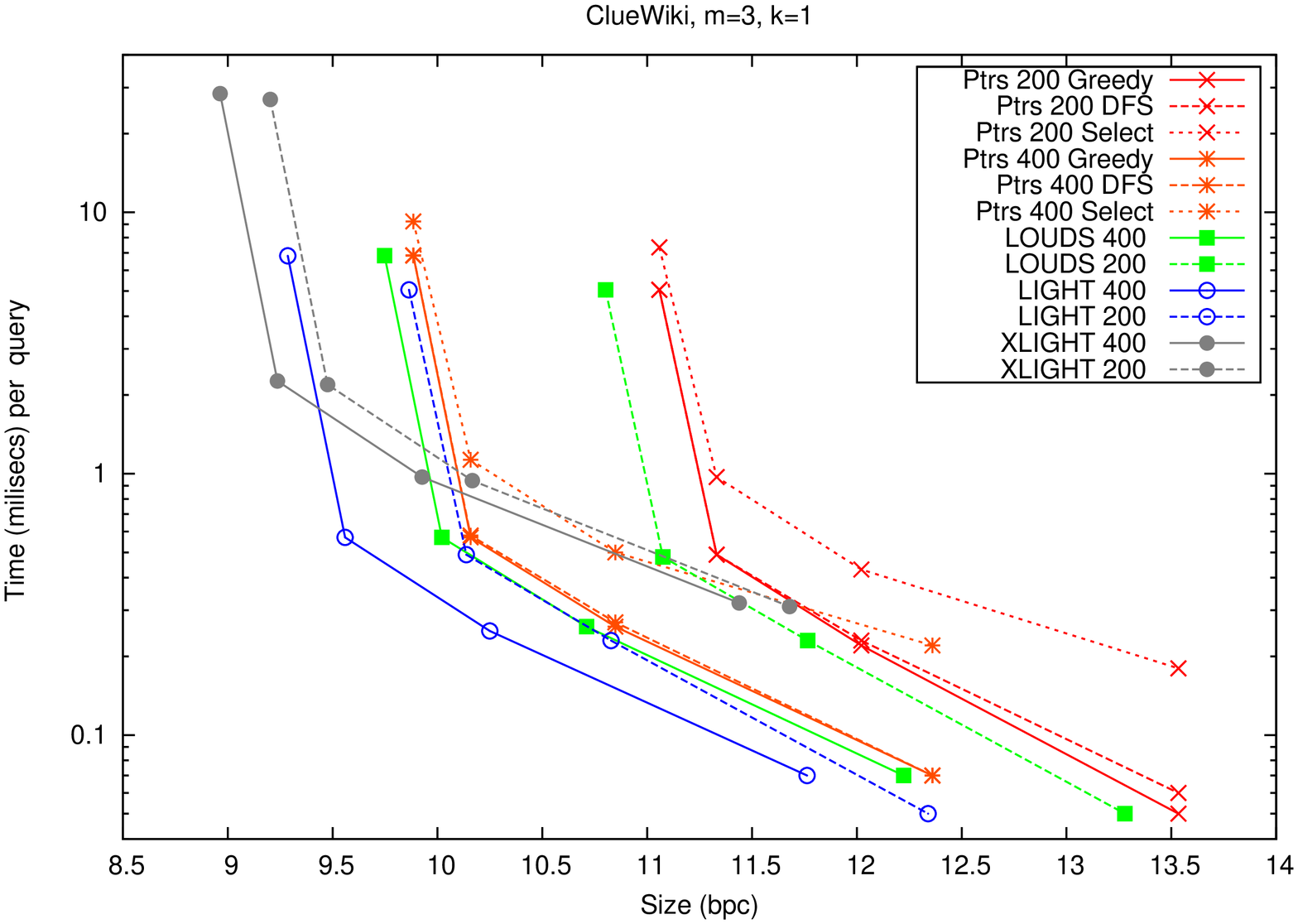}
\includegraphics[angle=-90,width=0.49\textwidth]{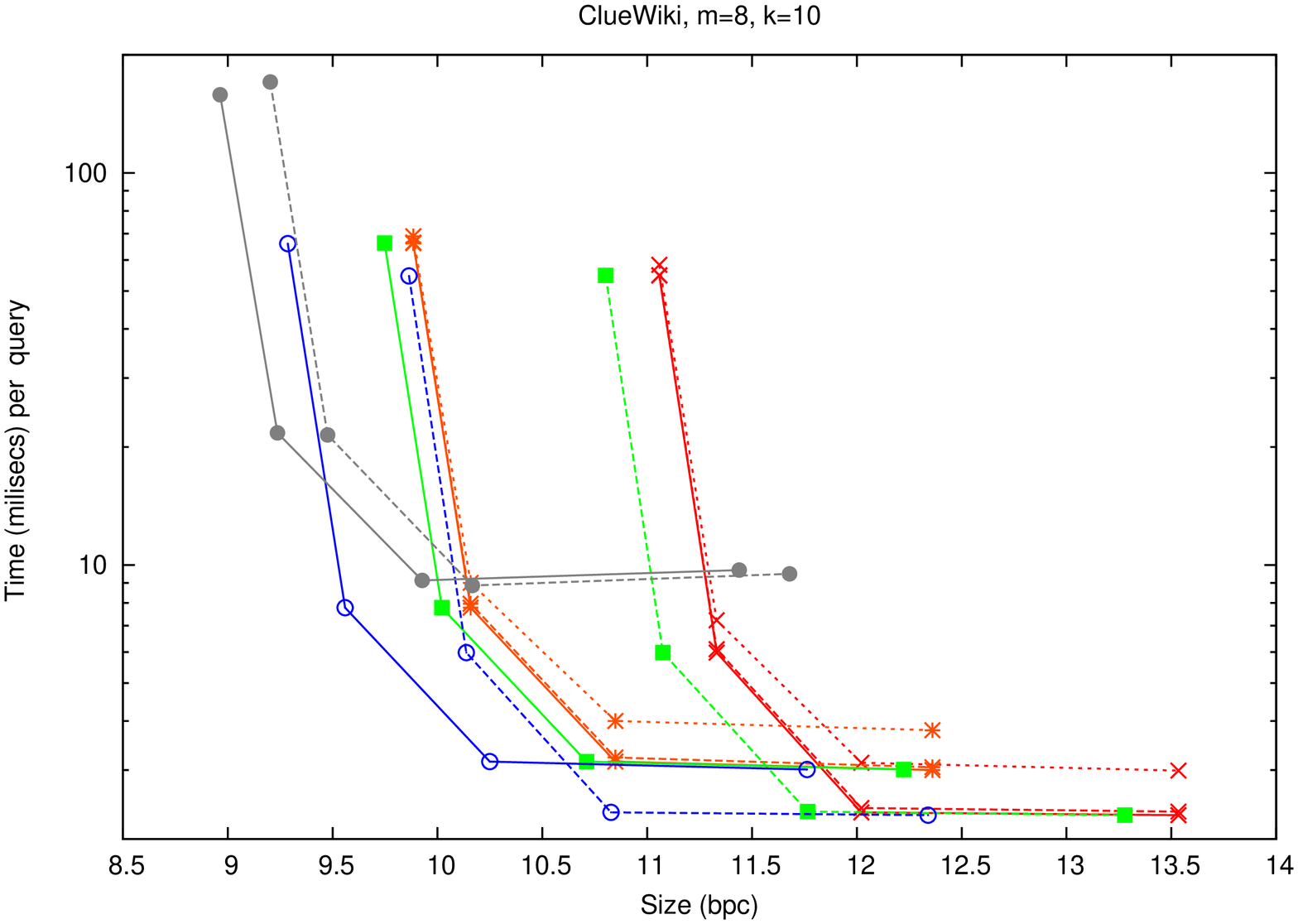}
}
\centerline{
\includegraphics[angle=-90,width=0.49\textwidth]{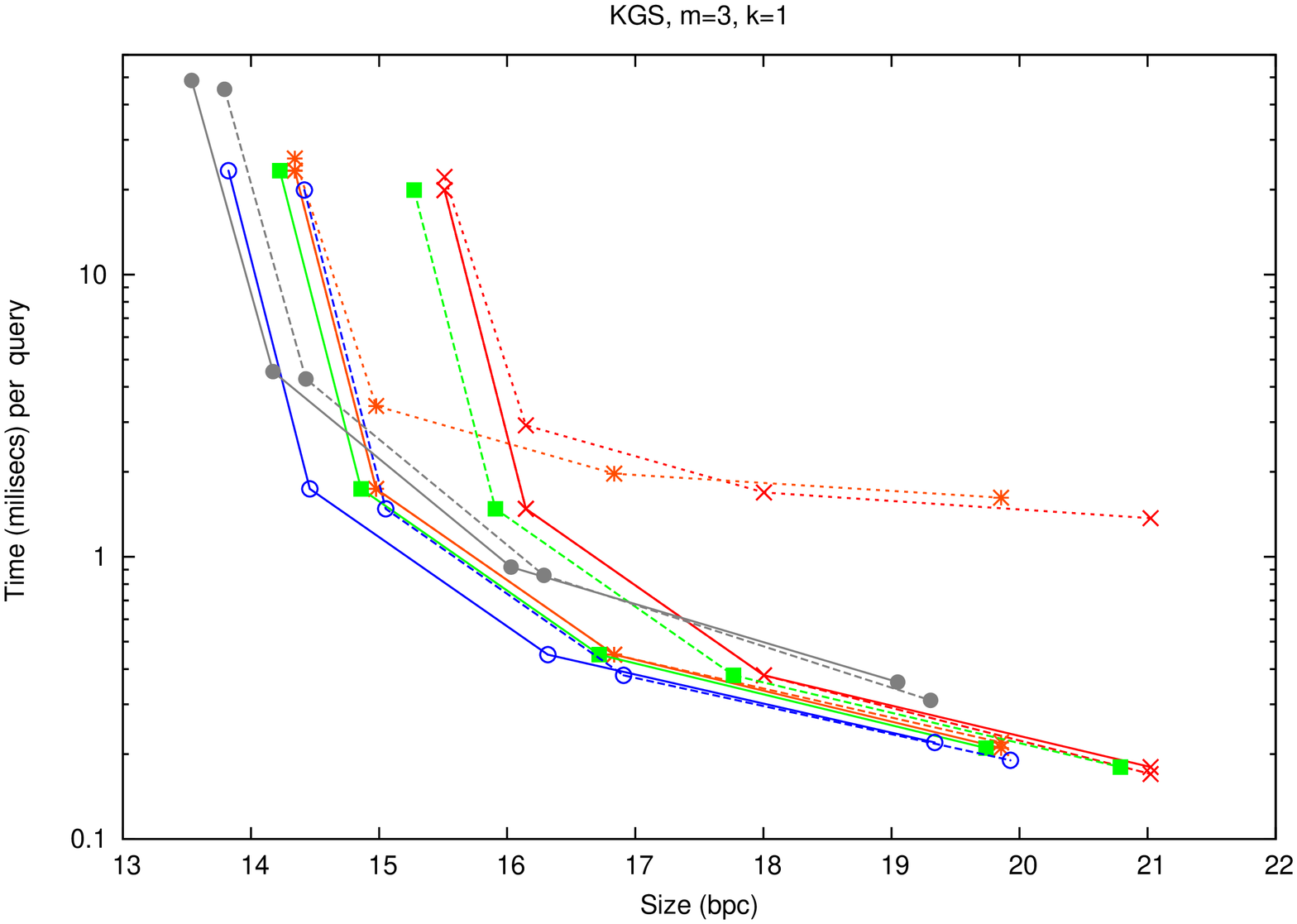}
\includegraphics[angle=-90,width=0.49\textwidth]{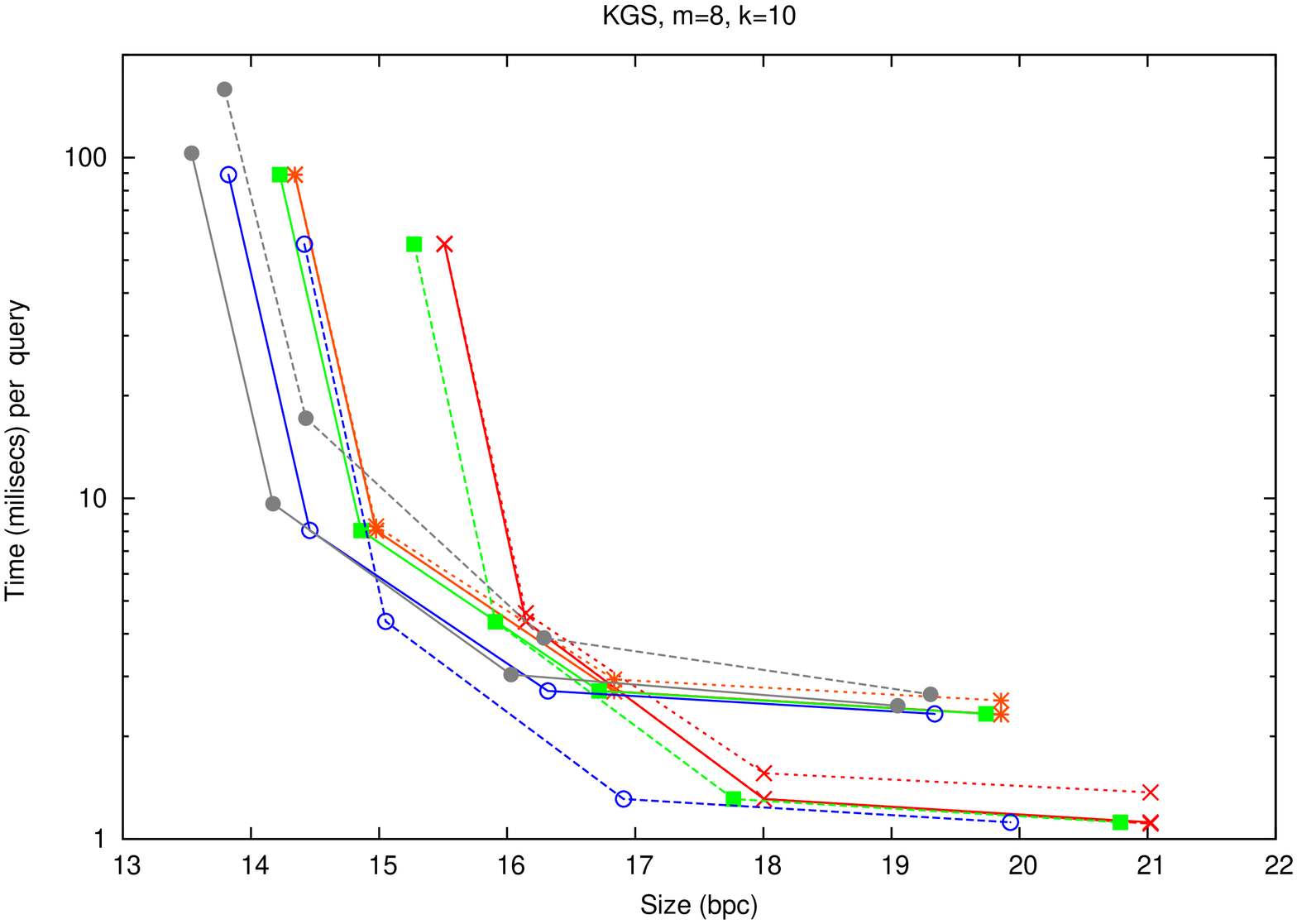}
}
\centerline{
\includegraphics[angle=-90,width=0.49\textwidth]{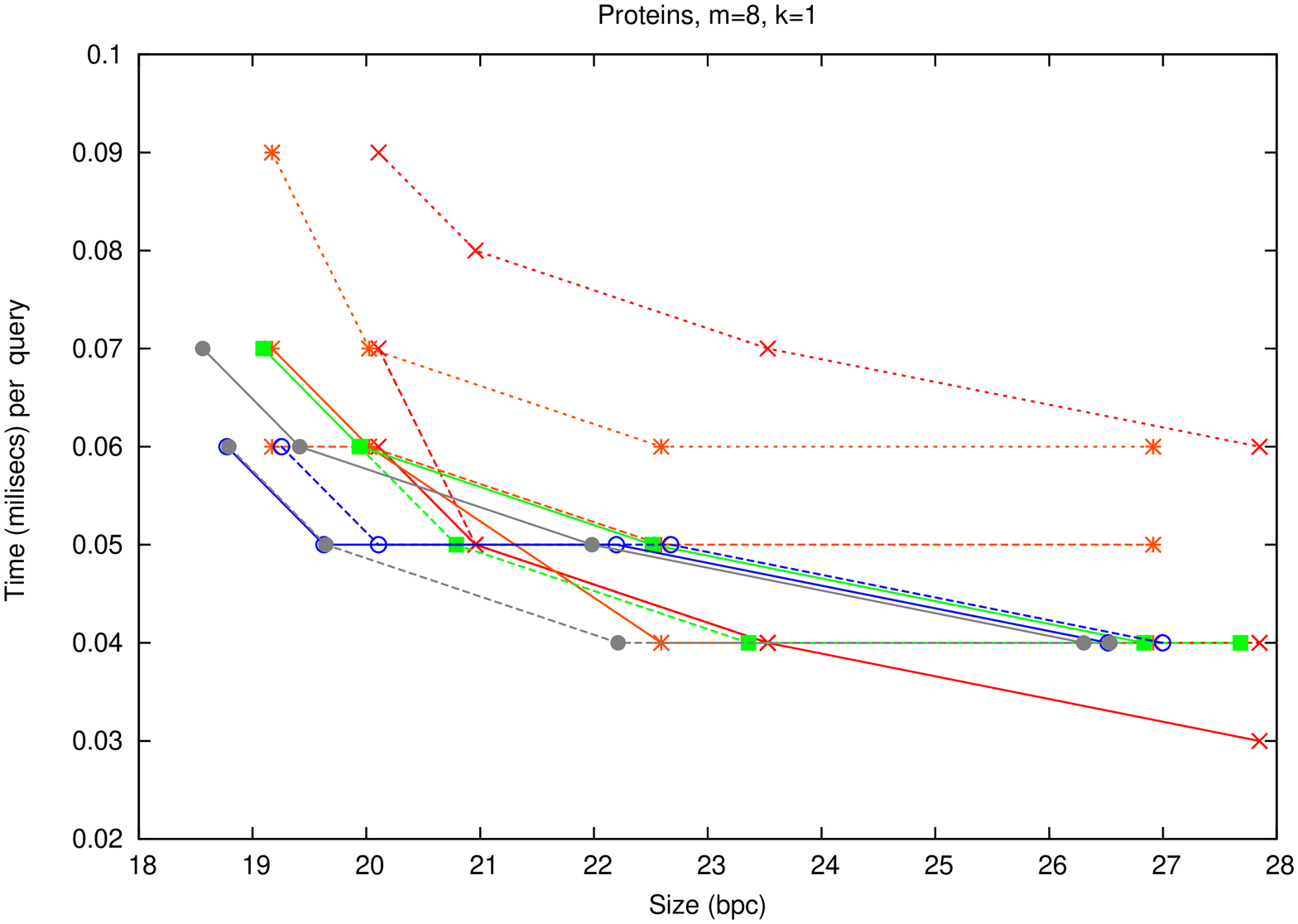}
\includegraphics[angle=-90,width=0.49\textwidth]{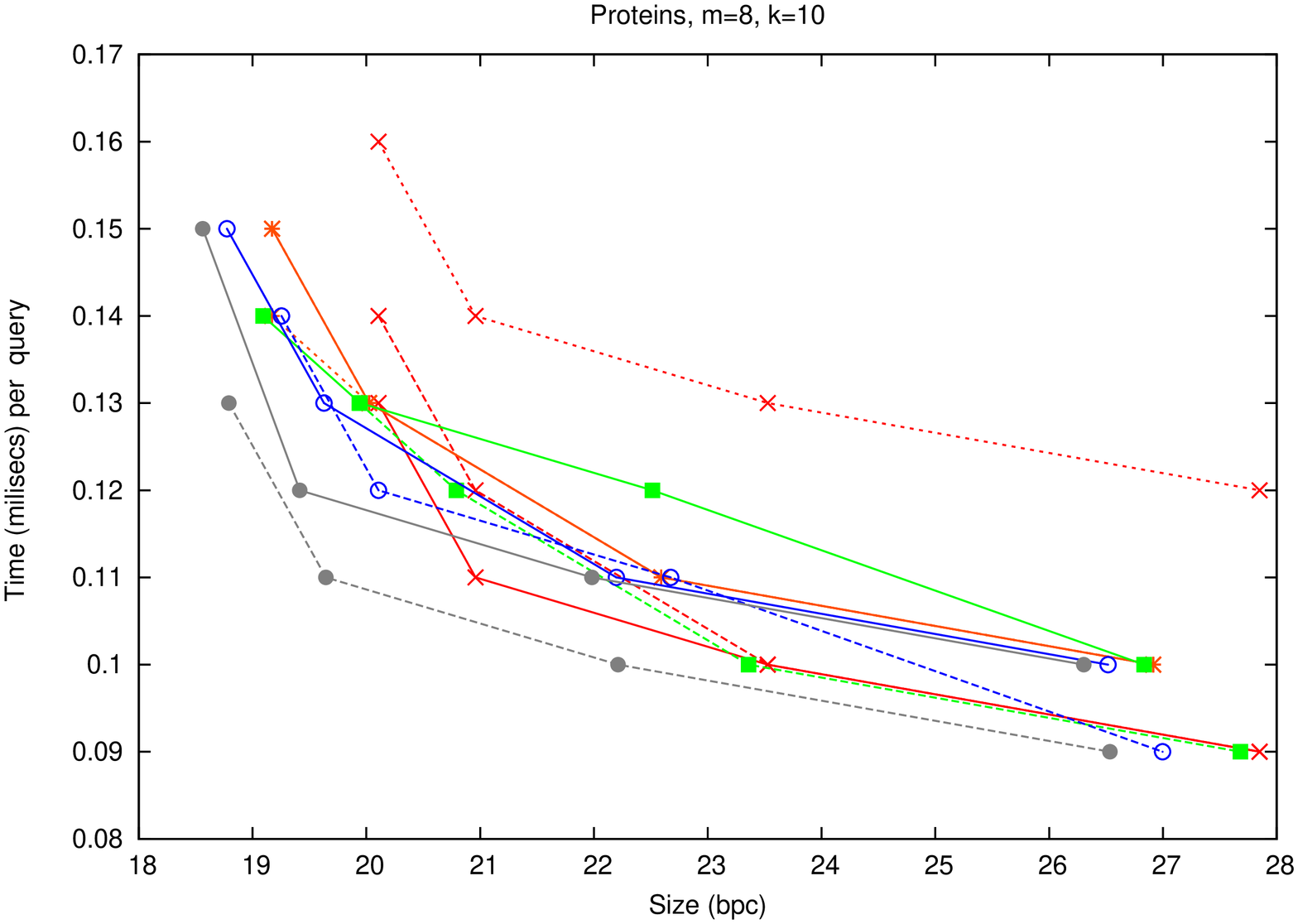}
}

\caption{Our different alternatives for top-$k$ queries. On the left for
$k=1$ and pattern length $m=3$; on the right for $k=10$ and $m=8$.}
\label{fig:prev}
\end{figure}

Figure~\ref{fig:prev} shows the results. Method {\tt Greedy} is always better
than {\tt Select} (up to 80\% better) and {\tt DFS} (up to 50\%), which
confirms intuition.
Using {\tt LOUDS} representation instead of {\tt Ptr} had almost no impact
on the time. This is because time needed to find the locus is usually
negligible compared with that to explore the uncovered leaves. Further
costless space gains are obtained with variant {\tt LIGHT}. Variant {\tt
XLIGHT}, instead, reduces the space of {\tt LIGHT} at a noticeable cost in
time that makes it not so interesting, except on {\bf Proteins}. In various 
cases the sparser sampling dominates the denser one, whereas in others the 
latter makes the structure faster if sufficient space is spent.

To compare with other techniques, we will use variant {\tt LIGHT} on
{\bf ClueWiki} and {\bf KGS}, and {\tt XLIGHT} on {\bf Proteins}, both with
$g'=400$. This combination will be called generically {\tt SSGST}.

\subsection{Comparison with Previous Work}

The second round of experiments compares ours with previous work.
The Greedy heuristic \cite{CNPT10} is run over different wavelet-tree
representations of the document array: a plain one ({\tt WT-Plain}) 
\cite{CNPT10}, a Re-Pair compressed one ({\tt WT-RP}), and a hybrid that 
at each wavelet tree level chooses between plain, Re-Pair, or entropy-based
compression of the bitmaps ({\tt WT-Alpha}) \cite{NPV11}. We combine these 
with our best implementation of Hon et al.'s structure (suffixing the previous 
names with {\tt +SSGST}). We also consider variant {\tt Goly+SSGST}
\cite{GNP10,HST11}, which runs the $rank$-based method ({\tt Select}) on top
of the fastest $rank$-capable sequence representation of the document
array (Golynski et al.'s \cite{GMR06}, which is faster than wavelet trees for 
$rank$ but does not support our more sophisticated algorithms; here we used 
the implementation at {\tt http://libcds.recoded.cl}).

Our new structures dominate most of the space-time map. When using
little space, variant {\tt WT-RP+SSGST} dominates, being only ocassionally
and slightly superseded by {\tt WT-RP}. When using more space, {\tt
WT-Alpha+SSGST} takes over, and finally, with even more space, {\tt
WT-Plain+SSGST} becomes the best choice. Most of the exceptions arise in
{\bf Proteins}, which due to its incompressibility \cite{NPV11} makes 
{\tt WT-Plain+SSGST} essentially the only interesting variant.
The alternative {\tt Goly+SSGST} is no case faster than a Greedy algorithm 
over plain wavelet trees ({\tt WT-Plain}), and takes more space.

\begin{figure}[p]
\centerline{
\includegraphics[angle=-90,width=0.49\textwidth]{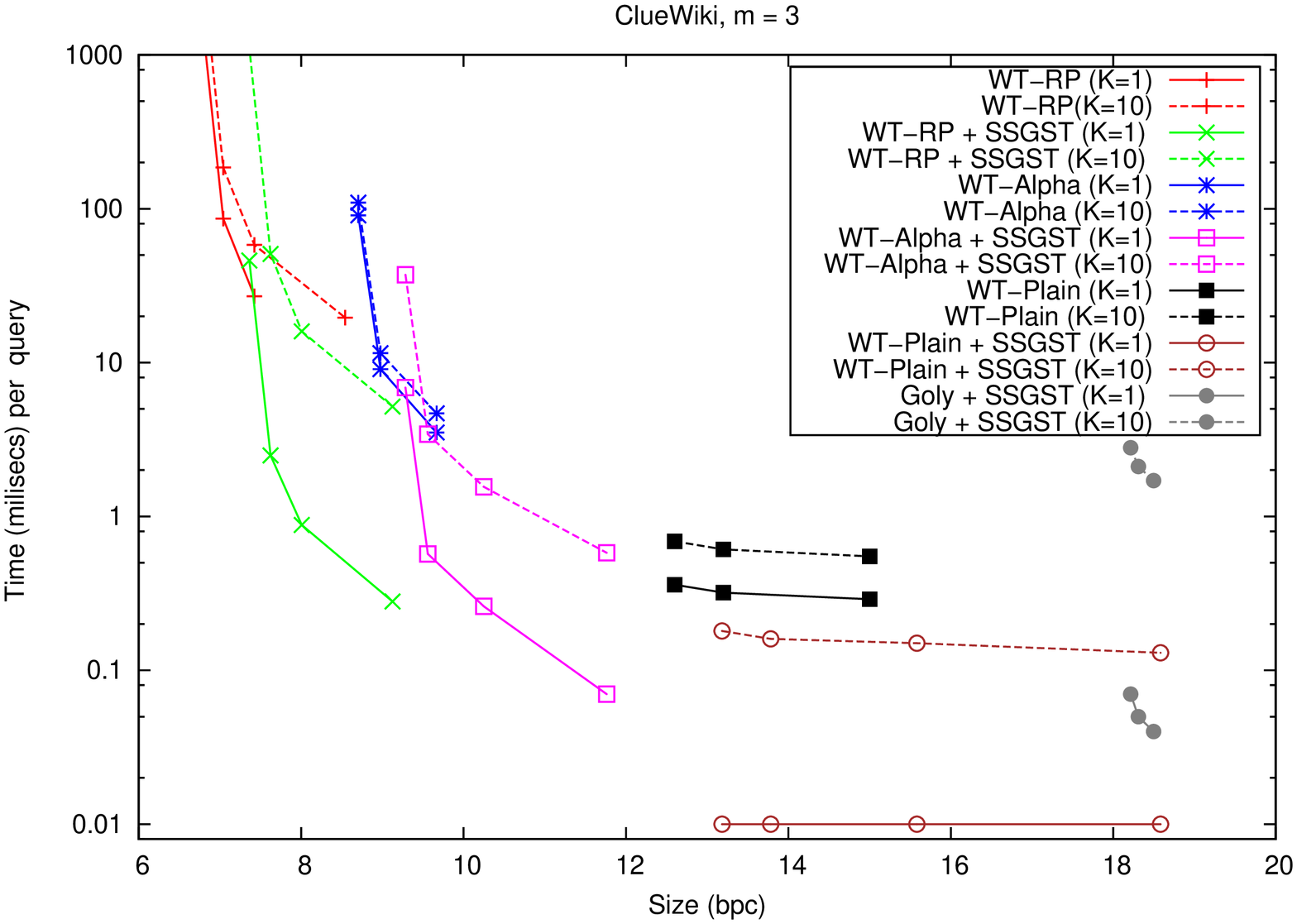}
\includegraphics[angle=-90,width=0.49\textwidth]{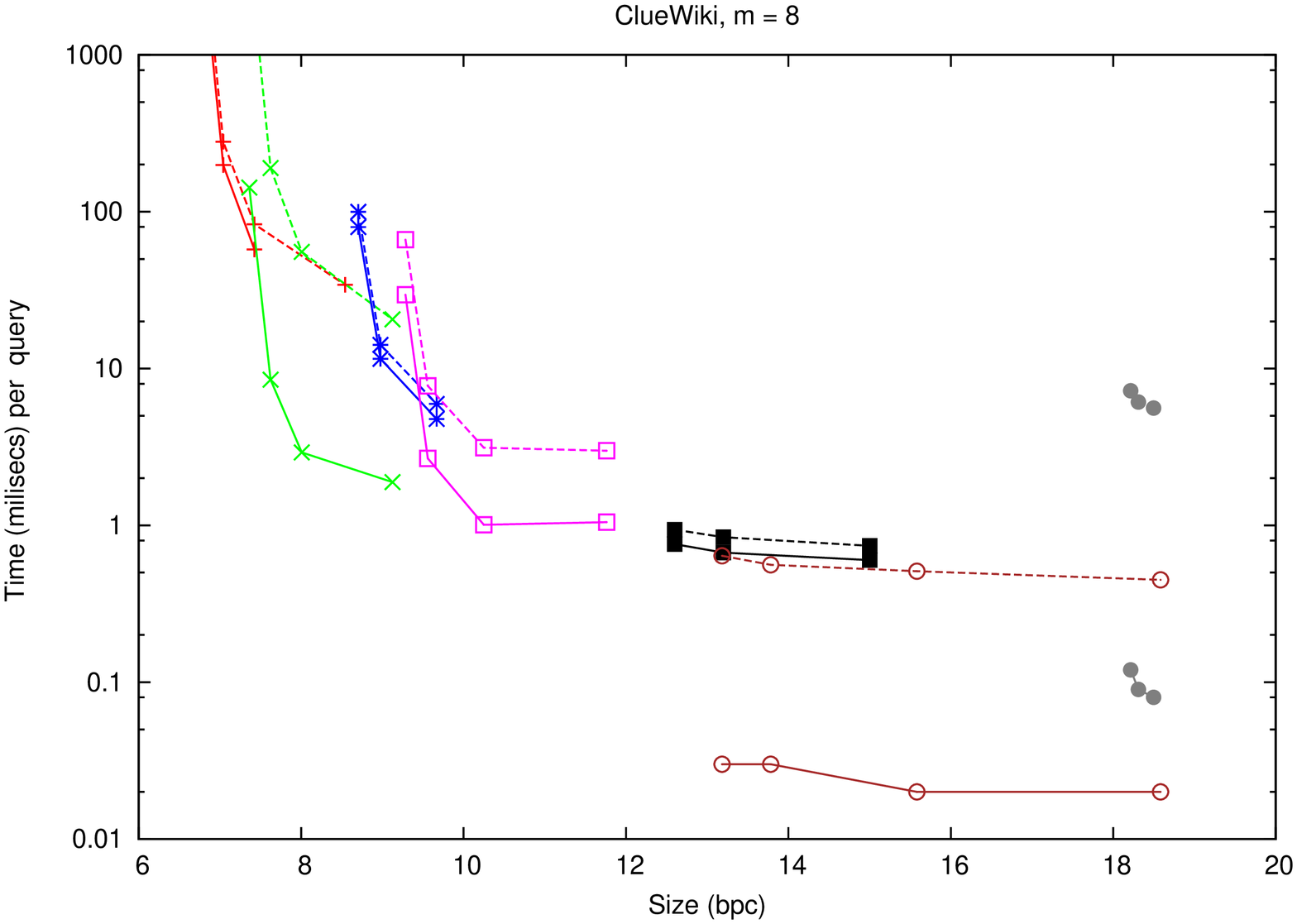}
}

\centerline{
\includegraphics[angle=-90,width=0.49\textwidth]{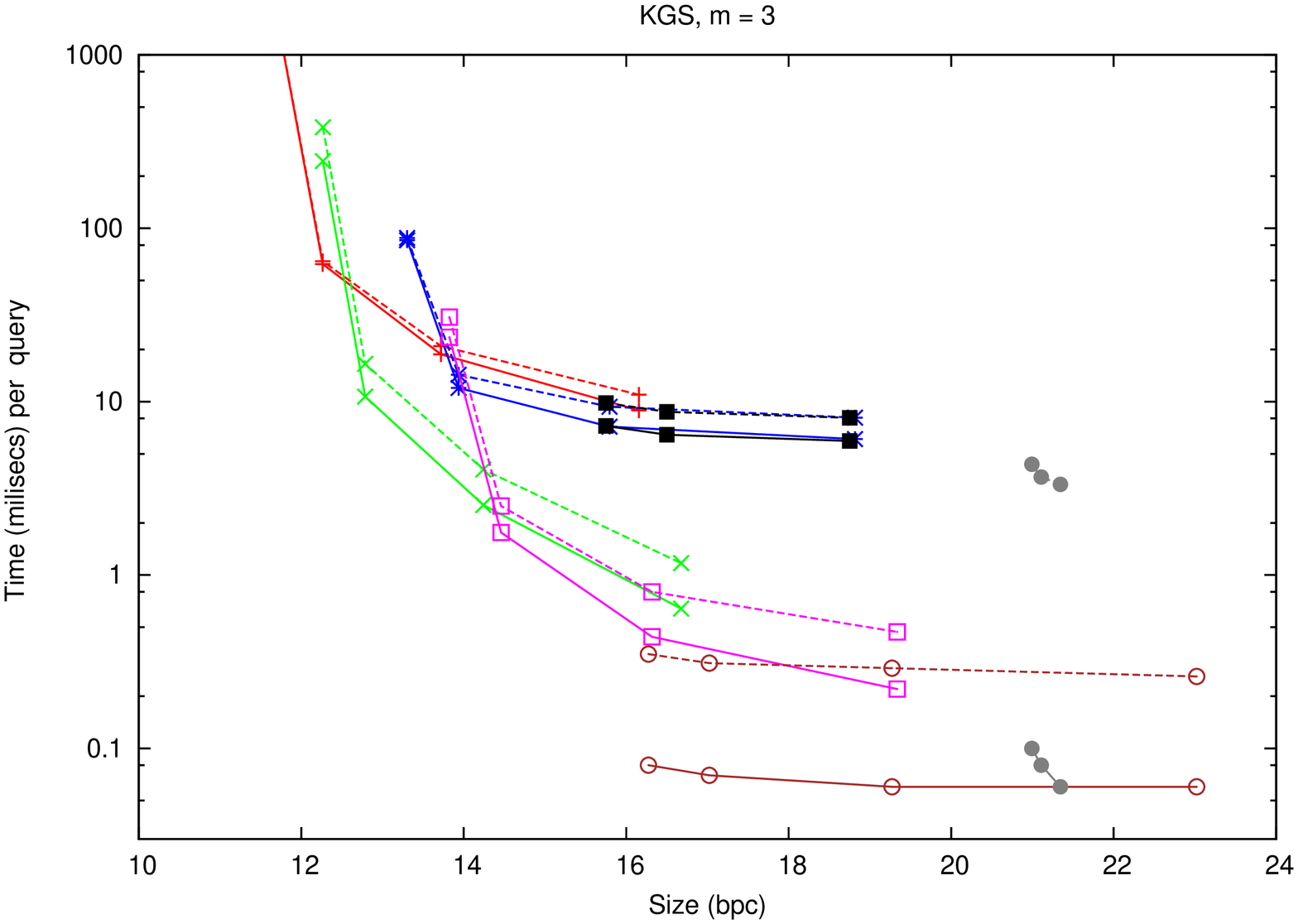}
\includegraphics[angle=-90,width=0.49\textwidth]{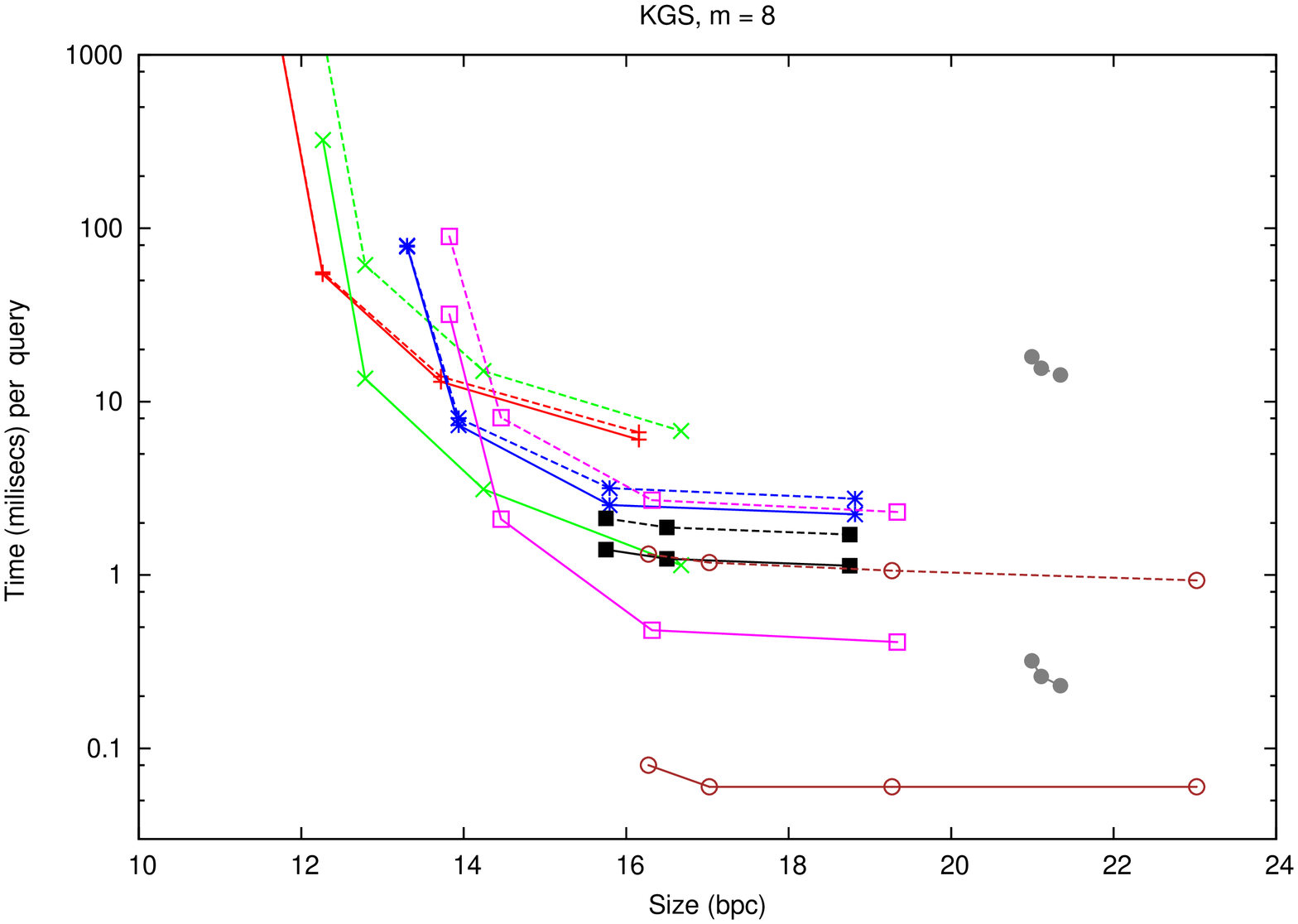}
}
\centerline{
\includegraphics[angle=-90,width=0.49\textwidth]{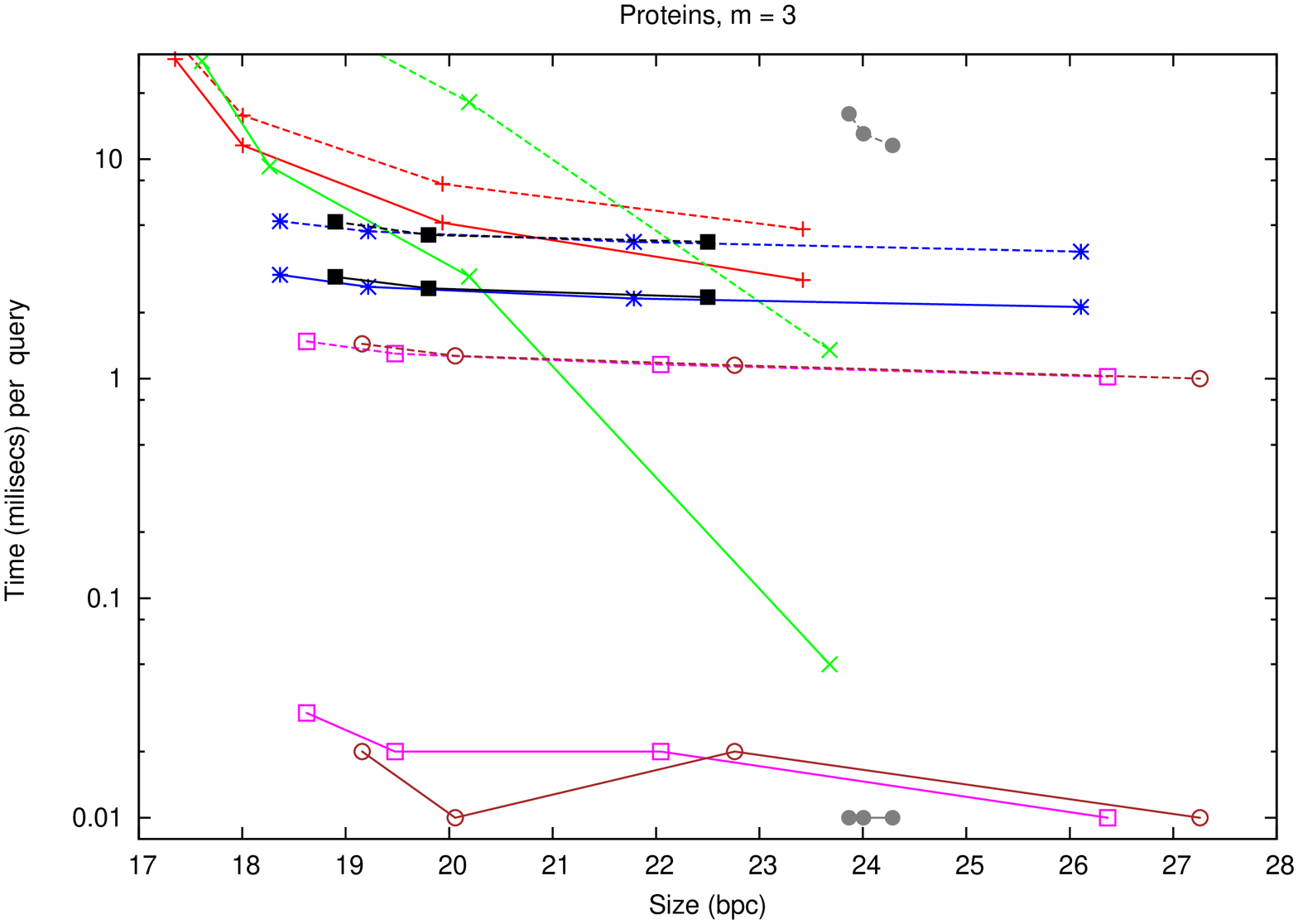}
\includegraphics[angle=-90,width=0.49\textwidth]{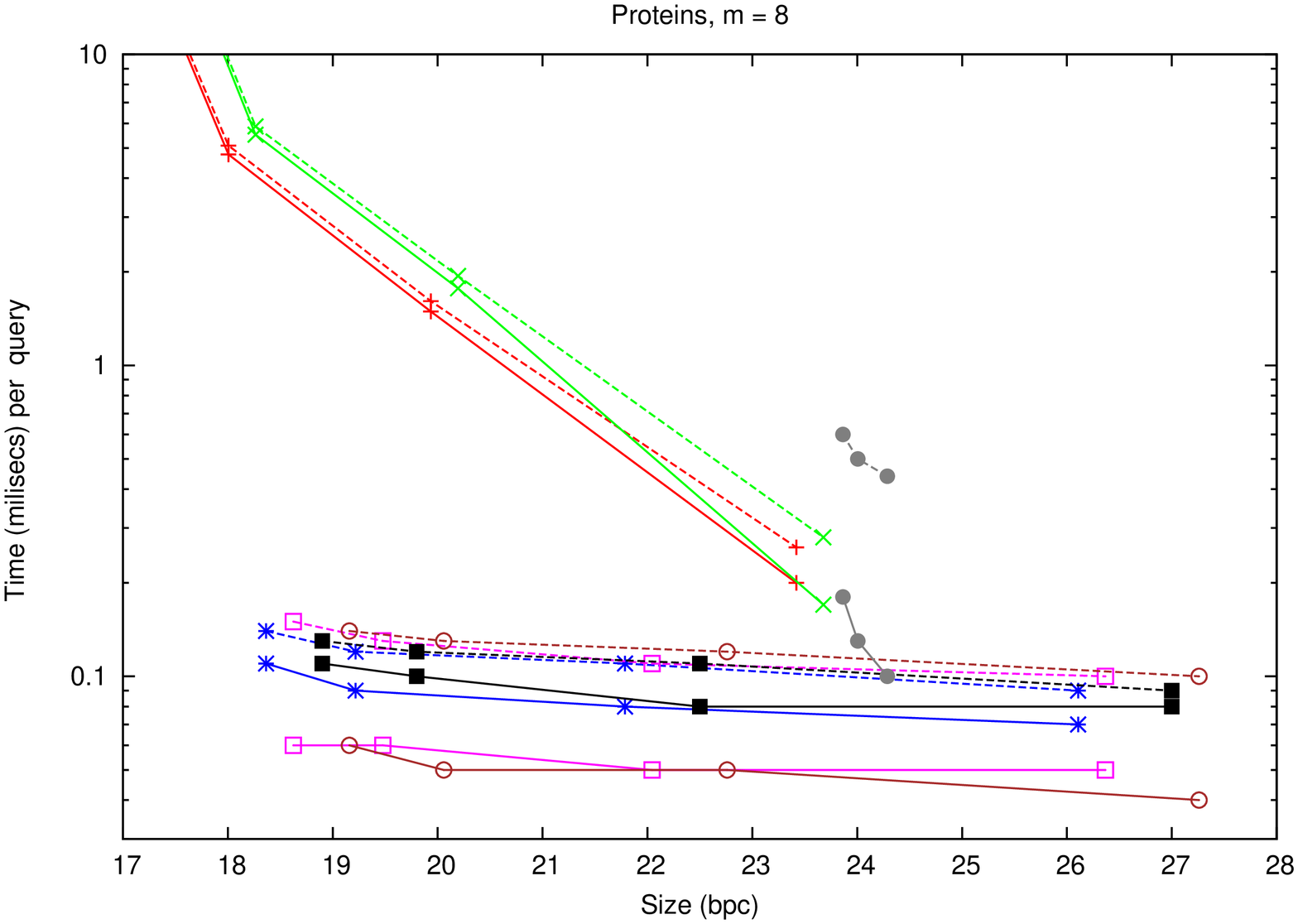}
}

\caption{Comparison with previous work. On the left, for $m=3$, and on the
right, for $m=8$.}
\label{fig:topk}
\end{figure}

\section{Future Work}

We can further reduce the space in exchange for possibly higher times. For
example the sequence of all precomputed top-$k$ candidates can be 
Huffman-compressed, as there is much repetition in the sets and a 
zero-order compression would yield space reductions of up to 25\% in
the case of {\bf Proteins}, the least compressible collection.
The pointers to those tables could also be removed, by separating the tables
by size, and computing the offset within each size using $rank$ on the sequence 
of classes of the nodes in $\tau$. Finally, values $[sp_v,ep_v]$ can be stored
as $[sp_v,ep_v-sp_v]$, using DACs for the second components \cite{BLN09}, as 
many such differences will be small. 

\newpage

\bibliographystyle{plain}
\bibliography{paper}

\end{document}